\documentclass[prd,a4paper]{revtex4}
\usepackage{epsfig}
\newcommand{\be}{\begin{equation}}
\newcommand{\ee}{\end{equation}}
\newcommand{\bea}{\begin{eqnarray}}
\newcommand{\eea}{\end{eqnarray}}
\newcommand{\bean}{\begin{eqnarray*}}
\newcommand{\eean}{\end{eqnarray*}}

\newcommand{\gapproxeq}{\lower
.7ex\hbox{$\;\stackrel{\textstyle >}{\sim}\;$}}
\newcommand{\lapproxeq}{\lower
.7ex\hbox{$\;\stackrel{\textstyle <}{\sim}\;$}}

\begin{document}

\bibliographystyle{unsrt}

\title{\bf Decays of $J/\psi$ and $\psi^\prime$ into
vector and pseudoscalar meson and the pseudoscalar
glueball-$q\bar{q}$ mixing }

\author{ Gang Li$^1$, Qiang Zhao$^{1,2}$}
\affiliation{1) Institute of High Energy Physics, Chinese Academy
of Sciences, Beijing, 100049, P.R. China}

\affiliation{2) Department of Physics, University of Surrey,
Guildford, GU2 7XH, United Kingdom}

\author{Chao-Hsi Chang$^{3,4}$}

\affiliation{3) CCAST (World Laboratory), P.O. Box 8730, Beijing
100080, P.R. China}

\affiliation{4) Institute of Theoretical Physics, Chinese Academy of
Sciences, Beijing, 100080, P.R. China}

\date{\today}

\begin{abstract}

We introduce a parametrization scheme for $J/\psi(\psi^\prime)\to
VP$ where the effects of SU(3) flavor symmetry breaking and doubly
OZI-rule violation (DOZI) can be parametrized by certain parameters
with explicit physical interpretations. This scheme can be used to
clarify the glueball-$q\bar{q}$ mixing within the pseudoscalar
mesons. We also include the contributions from the electromagnetic
(EM) decays of $J/\psi$ and $\psi^\prime$ via
$J/\psi(\psi^\prime)\to \gamma^*\to VP$. Via study of the isospin
violated channels, such as $J/\psi(\psi^\prime)\to \rho\eta$,
$\rho\eta^\prime$, $\omega\pi^0$ and $\phi\pi^0$, reasonable
constraints on the EM decay contributions are obtained. With the
up-to-date experimental data for $J/\psi(\psi^\prime)\to VP$,
$J/\psi(\psi^\prime)\to \gamma P$ and $P\to \gamma\gamma$, etc, we
arrive at a consistent description of the mentioned processes with a
minimal set of parameters. As a consequence, we find that there
exists an overall suppression of the $\psi^\prime\to 3g$ form
factors, which sheds some light on the long-standing ``$\rho\pi$
puzzle". By determining the glueball components inside the
pseudoscalar $\eta$ and $\eta^\prime$ in three different
glueball-$q\bar{q}$ mixing schemes, we deduce that the lowest
pseudoscalar glueball, if exists, has rather small $q\bar{q}$
component, and it makes the $\eta(1405)$ a preferable candidate for
$0^{-+}$ glueball.

\end{abstract}

\maketitle






\vspace{1cm}

\section{Introduction}

Charmonium decays into light hadrons provides unique places to probe
light hadron structures. In particular, in the hadronic decays of
charmonia such as $J/\psi$, $\eta_c$ and $\chi_{cJ}$ etc, the
annihilation of the heavy $c\bar{c}$ pair into intermediate gluons,
which must be then hadronized into hadrons, could favor the
production of unconventional hadrons such as glueball and hybrids in
the final-state. Such states, different from the conventional
$q\bar{q}$ or $qqq$ structures in the non-relativistic constituent
quark model, can serve as a direct test of QCD as a non-Abelian
gauge theory.

In the past decades, the exclusive hadronic decays of
$J/\psi(\psi^\prime)\to VP$ have attracted a lot of attention in
both experiment and theory. These processes, in which the helicity
is non-conserved, must be suppressed according to the selection
rule of pQCD hadronic helicity conservation due to the vector
nature of gluons~\cite{brodsky-lepage-81}. According to the ``pQCD
power" suppression of helicity conservation and breaking, one
should have $BR(\psi^\prime\to\rho\pi)/BR(J/\psi\to\rho\pi)\simeq
(M_{J/\psi}/M_{\psi^\prime})^6\sim 0.35$~\cite{brodsky-lepage-81}.
However, the experimental data show that this ratio is badly
violated in reality,
$BR(\psi^\prime\to\rho\pi)/BR(J/\psi\to\rho\pi)\simeq (0.2\pm
0.1)\%$~\cite{pdg2006}, i.e. much greatly suppressed. It led to
the so-called ``$\rho\pi$ puzzle" in the study of $J/\psi$ and
$\psi^\prime$ exclusive decays, and initiated a lot of interests
to the relevant
issues~\cite{hou-soni,karl-roberts,pinsky-85,brodsky-lepage-tuan,ct,pinsky-90,brodsky-karliner,li-bugg-zou,chen-braaten,gerald,feldmann,suzuki,rosner,wym,liu-zeng-li}.
An alternative expression of the ``$\rho\pi$ puzzle" is via the
ratios between $J/\psi$ and $\psi^\prime$ annihilating into three
gluons and a single photon: \bea R &\equiv &
\frac{BR(\psi^\prime\to hadrons)}{BR(J/\psi\to
hadrons)}\nonumber\\
&\simeq &\frac{BR(\psi^\prime\to e^+ e^-)}{BR(J/\psi\to e^+
e^-)}\simeq 12\% ,
\eea
which is empirically called ``12\% rule". Since most of those
exclusive decays for $J/\psi$ and $\psi^\prime$ seem to abide by
this empirical rule reasonably well, it is puzzling that the
ratios for $\rho\pi$ and $K^*\bar{K}+c.c.$ deviate dramatically
from it. A recent article by Mo, Yuan and Wang provides a detailed
review of the present available explanations (see Ref.
~\cite{yuan-review} and references therein).

A catch-up of this subject is an analysis by authors here for the
electromagnetic decays of $J/\psi(\psi^\prime)\to\gamma^*\to
VP$~\cite{zhao-li-chang}. There, it is shown that although the EM
contributions to $J/\psi \to VP$ are generally small relative to
the strong decays as found by many other studies~\cite{seiden},
they may turn to be more competitive in $\psi^\prime\to VP$
compared with $\psi^\prime\to 3g \to VP$ amplitudes, and produce
crucial interferences~\cite{zhao-li-chang}. For a better
understanding of the ``$\rho\pi$ puzzle", a thorough study of
$J/\psi(\psi^\prime)\to VP$ including both strong and EM
transitions and accommodating the up-to-date experimental
information~\cite{bes-05,bes-04,bes-04b,bes-05b} should be
necessary.

The reaction channels $J/\psi(\psi^\prime)\to VP$ also give access
to another interesting issue in non-perturbative QCD. They may be
used to probe the structure of the isoscalar $0^{-+}$, i.e. $\eta$
and $\eta^\prime$ in their recoiling isoscalar vector mesons
$\omega$ or $\phi$. Since vector meson $\omega$ and $\phi$ are
almost ideally mixed, i.e. $\omega=(u\bar{u}+d\bar{d})/\sqrt{2}$ and
$\phi=s\bar{s}$, the decay of $J/\psi(\psi^\prime)\to \omega \eta$,
$\omega\eta^\prime$, $\phi\eta$, and $\phi\eta^\prime$ will provide
information about $\eta$ and $\eta^\prime$, which can be produced
via the so-called singly OZI disconnected (SOZI) processes for the
same flavor components, and doubly OZI disconnected (DOZI) processes
for different flavor components.

Nonetheless, it can also put experimental constraints on the
octet-singlet mixing as a consequence of the $U(1)_A$ anomaly of
QCD~\cite{fritzsch-jackson,isgur-76,KO,ktchao,SSW,witten,gilman-kauffman,Leutwyler,fks}.
Recently, Leutwyler~\cite{Leutwyler}, and Feldmann and his
collaborators~\cite{fks} propose a $\eta$-$\eta^\prime$ mixing
scheme in the quark flavor basis with an assumption that the decay
constants follow the pattern of particle state mixing, and thus
are controlled by specific Fock state wavefunctions at zero
separation of the quarks, while the state mixing is referred to
the mixing in the overall wavefunctions. In the quark flavor basis
it is shown that higher Fock states (specifically, $|gg\rangle$)
due to the anomaly give rise to a relation between the mixing
angle and decay constants. This analysis brings the question of
gluon components inside $\eta$ and $\eta^\prime$ to much closer
attention from both experiment and theory~\cite{fks,kroll-03}, and
initiates a lot of interests in the pseudoscalar sector, in
particular, in line with the search for pseudoscalar glueball
candidates.

The lattice QCD (LQCD) calculations predict a mass for the lowest
pseudoscalar glueball around 2.5 GeV~\cite{lqcd}, which is higher
than a number of $0^{-+}$ resonances observed in the range of
1$\sim$ 2.3 GeV. However, since the present LQCD calculations are
based on quenched approximation, the glueball spectrum is still an
open question in theory. In contrast, QCD phenomenological
studies~\cite{cfl,farrar,cakir-farrar,faddeev-04} favor a much lower
$0^{-+}$ glueball mass, and make $\eta(1405)$ a good glueball
candidate due to its strong couplings to $K\bar{K}\pi$ and
$a_0(980)\pi$ and absence in $\gamma\gamma\to K\bar{K}\pi$ and
$\eta\pi\pi$~\cite{pdg2006}. Study of $J/\psi\to VP$ in flavor
parametrization schemes was pursued in Ref.~\cite{seiden}, where the
$\eta$ and $\eta^\prime$ were treated as eigenstates of $q\bar{q}$.
By assuming the $\eta(1405)$ to be a glueball essentially and
introducing a DOZI suppression factor for $\phi G$ and $\omega G$
amplitudes relative to the SOZI processes, the authors seemed to
have underestimated the branching ratios for $BR(J/\psi\to
\phi\eta(1405))$ and $BR(J/\psi\to \omega\eta(1405))$. In
Ref.~\cite{li-yu-fang} a mixing scheme for $\eta$, $\eta^\prime$ and
$\eta(1405)$ in a basis of $(u\bar{u}+d\bar{d})/\sqrt{2}$,
$s\bar{s}$ and $G$ (glueball) was proposed and studied in $J/\psi\to
VP$. $\eta(1405)$ was found to be dominated by the glueball
component, while the glueball component in $\eta^\prime$ was also
sizeable. In both studies, the EM contributions were included as a
free parameter in $J/\psi\to VP$. However, as we now know that the
EM contributions in $\psi^\prime$ are
important~\cite{zhao-li-chang}. Thus, a coherent study of
$J/\psi(\psi^\prime)\to VP$ including the EM contributions would be
ideal for probing the structure of the pseudoscalar mesons.

On the theoretical side there is only one possible term in the
effective Lagrangian for $J/\psi(\psi^\prime)\to VP$ and it will
reduce the number of the parameters needed in phenomenological
studies. On the experimental side many new data with high accuracy
are available. Hence, in this work, we shall revisit
$J/\psi(\psi^\prime)\to VP$ trying to clarify the following
points: (i) What is the role played by the EM decay transitions?
By isolating and constraining the EM transitions in a VMD model,
we propose a parametrization scheme for the strong transitions
where the SU(3) flavor symmetry breaking and DOZI violation
effects can be accommodated. A reliable calculation of the EM
transitions in turn will put a reasonable constraint on the strong
decay transitions in $J/\psi(\psi^\prime)\to VP$, and thus
mechanisms leading to the ``$\rho\pi$ puzzle" can be highlighted.
(ii) We shall probe the glueball components within $\eta$ and
$\eta^\prime$ based on the available experimental data and
different quarkonia-glueball mixing schemes, and predict the
$0^{-+}$ glueball production rate in $J/\psi(\psi^\prime)$
hadronic decays.

As follows, we first present the parametrization scheme for
$J/\psi(\psi^\prime)\to VP$ and introduce the glueball components
into the pseudoscalars. We then briefly discuss the VMD model for
$J/\psi(\psi^\prime)\to\gamma^*\to VP$. In Section III, we present
the model calculation results with detailed discussions. A summary
will be given in Section IV.

\section{Model for $J/\psi(\psi^\prime)\to VP$}

In this Section, we first introduce a simple rule to parametrize
out the strong decays of $J/\psi(\psi^\prime)\to VP$. We then
introduce a VMD model for the $J/\psi(\psi^\prime)$ EM decays into
$VP$. A recent study of $J/\psi(\psi^\prime)\to\gamma^*\to VP$
shows that the EM decay contributions become important in
$\psi^\prime$ decays due to its large partial decay width to $e^+
e^-$ though their importance is not so significant in $J/\psi\to
VP$~\cite{zhao-li-chang}. In the isoscalar pseudoscalar sector, we
introduce three different glueball-$q\bar{q}$ mixing schemes for
$\eta$, $\eta^\prime$ and a glueball candidate
$\eta^{\prime\prime}$. Its production in $J/\psi(\psi^\prime)$
hadronic decays can then be factorized out and estimated.

\subsection{Decay of $J/\psi(\psi^\prime) \to VP$ via
strong interaction}

The strong decay of $J/\psi (\psi^\prime)\to VP$ via $3g$ can be
factorized out in a way similar to Ref.~\cite{zhao-chic}.

First, we define the strength of the non-strange singly OZI
disconnected (SOZI) process as
\be
g_{J/\psi(\psi^\prime)}\equiv \langle (q\bar{q})_V (q\bar{q})_P
|V_0|J/\psi(\psi^\prime)\rangle,
\ee
where $V_0$ denotes the $3g$ decay potential of the charmonia into
two non-strange $q\bar{q}$ pairs of vector and pseudoscalar via
SOZI processes. But it should be noted that the subscription $V$
and $P$ do not mean that the quark-antiquark pairs are flavor
eigenstates of vector and pseudoscalar mesons. The amplitude
$g_{J/\psi(\psi^\prime)}$ is proportional to the charmonium
wavefunctions at origin. Thus, it may have different values for
$J/\psi$ and $\psi^\prime$.

In order to include the SU(3) flavor symmetry breaking effects in
the transition, we introduce \be R \equiv  \langle
(q\bar{s})_V(s\bar{q})_P|V_0|J/\psi(\psi^\prime)\rangle /
g_{J/\psi(\psi^\prime)} = \langle
(s\bar{q})_V(q\bar{s})_P|V_0|J/\psi(\psi^\prime)\rangle /
g_{J/\psi(\psi^\prime)}, \ee which implies the occurrence of the
SU(3) flavour symmetry breaking at each vertex where a pair of
$s\bar{s}$ is produced, and $R=1$ is in the SU(3) flavour symmetry
limit. For the production of two $s\bar{s}$ pairs via the SOZI
potential, the recognition of the SU(3) flavor symmetry breaking in
the transition is accordingly
\be R^2 \simeq \langle
(s\bar{s})_V(s\bar{s})_P|V_0|J/\psi(\psi^\prime)\rangle /
g_{J/\psi(\psi^\prime)} \ . \ee

Similar to Ref.~\cite{zhao-chic}, the DOZI process is
distinguished from the SOZI ones by the gluon counting rule. A
parameter $r$ is introduced to describe the relative strength
between the DOZI and SOZI transition amplitudes:
\be
r\equiv \langle
(s\bar{s})_V(q\bar{q})_P|V_1|J/\psi(\psi^\prime)\rangle /
g_{J/\psi(\psi^\prime)} = \langle
(q\bar{q})_V(s\bar{s})_P|V_1|J/\psi(\psi^\prime)\rangle /
g_{J/\psi(\psi^\prime)},
\ee
where $V_1$ denotes the charmonium decay potential via the DOZI
processes; In the circumstance that the OZI rule is respected, one
expects $|r|\sim 0$.

Through the above definitions, we express the amplitudes for
$J/\psi(\psi^\prime)\to V(q\bar{q})_P$ as
\begin{eqnarray}
M_{K^{* +}K^-} &=& \langle K^{* +}K^-|V_0 |J/\psi(\psi^\prime)\rangle = g_{J/\psi(\psi^\prime)} R \ , \\
M_{\rho^+\pi^-} &=& \langle \rho^+\pi^-|V_0 |J/\psi(\psi^\prime)\rangle = g_{J/\psi(\psi^\prime)} \ , \\
M_{\rho^0\pi^0} &=& \langle \rho^0\pi^0|V_0
|J/\psi(\psi^\prime)\rangle  =  g_{J/\psi(\psi^\prime)} \ , \\
M_{\phi (s\bar {s})}&=& \langle\phi (s\bar{s})_P
|V_0+V_1|J/\psi(\psi^\prime) \rangle  = R^2g_{J/\psi(\psi^\prime)}(1+r),\\
 M_{\phi (n\bar {n})}&=& \langle \phi (n\bar{n})_P
 |V_1 |J/\psi (\psi^\prime)\rangle = \sqrt {2}r  \langle \phi(s\bar{s})_P
 |V_0 |J/\psi(\psi^\prime)\rangle = \sqrt {2} rRg_{J/\psi(\psi^\prime)}, \\
M_{\omega (s\bar{s})}&=& \langle \omega
(s\bar{s})_P|V_1 |J/\psi(\psi^\prime) \rangle  = \sqrt {2}R r g_{J/\psi(\psi^\prime)} ,\\
 M_{\omega (n\bar {n})}&=& \langle \omega (n\bar{n})_P
 |V_0 + V_1 |J/\psi(\psi^\prime) \rangle =  g_{J/\psi(\psi^\prime)}
(1+2r), \label{amplitude1}
\end{eqnarray}
where the channels $K\bar{K^*}$ and $\rho^-\pi^+$ have the same
expressions as their conjugate channels; $n\bar{n}\equiv
(u\bar{u}+d\bar{d})/\sqrt{2}$ is the non-strange isospin singlet.
We have assumed that $\phi$ and $\omega$ are ideally mixed and
they are pure $s\bar{s}$ and $n\bar{n}$, respectively.

The recoiled pseudoscalars by $\phi$ and $\omega$ can be either
$\eta$ or $\eta^\prime$ for which we consider that a small
glueball component will mix with the dominant $q\bar{q}$ in the
quark flavour basis. The detailed discussion about the $q\bar{q}$
and glueball mixings in $\eta$ and $\eta^\prime$ will be given in
the next section. Here, we relate the production of glueball $G$
in $J/\psi(\psi^\prime)\to \omega G$ and $\phi G$ with the basic
amplitude $g_{J/\psi(\psi^\prime)}$ by assuming the validity of
gluon counting rule in the
transitions~\cite{zhao-chic,close-zhao-scalar}, i.e.
\be
\langle (q\bar{q})_V G| V_3 |J/\psi(\psi^\prime)\rangle \equiv
g_{J/\psi(\psi^\prime)} \ ,
\ee
where $V_3$ denotes the glueball $G$ production potential
recoiling a flavour singlet $q\bar{q}$. This can be regarded as
reasonable since generally the glueball does not pay a price for
its couplings to gluons. Consequently, we have
\bea
M_{\phi G} &=& \langle \phi G| V_3 |J/\psi(\psi^\prime)\rangle =
g_{J/\psi(\psi^\prime)} R
\nonumber\\
M_{\omega G} &=& \langle \omega G| V_3 |J/\psi(\psi^\prime)\rangle
= \sqrt{2} g_{J/\psi(\psi^\prime)} \ ,
\eea
with the SU(3) flavor symmetry breaking considered for the $\phi$
production.

The above parametrization highlights several interesting features
in those decay channels. It shows that the decay of
$J/\psi(\psi^\prime) \to \rho\pi$ is free of interferences from
the DOZI processes and possible SU(3) flavor symmetry breaking.
Ideally, such a process will be useful for us to determine
$g_{J/\psi(\psi^\prime)}$. The decay of $J/\psi(\psi^\prime) \to
K^* \bar{K}+c.c.$ is also free of DOZI interferences, but
correlates with the SU(3) breaking. These two sets of decay
channels will, in principle, allow us to determine the basic decay
amplitude $g_{J/\psi(\psi^\prime)}$ and the SU(3) flavor breaking
effects, which can then be tested in other channels. For
$\phi\eta$, $\phi\eta^\prime$, $\omega\eta$ and
$\omega\eta^\prime$, the transition amplitudes will also depend on
the $\eta$-$\eta^\prime$ mixing angles and we will present
detailed discussions in later part.

In the calculation of the partial decay width, we must apply the
commonly used form factor
\begin{eqnarray}
{\cal {F}}^2({\bf P}) \equiv |{\bf P}|^{2l}\exp ({-{\bf
P}^2/{8\beta^2}}),
\end{eqnarray}
where ${\bf P}$ and the $l$ are the three momentum and the
relative orbit angular momentum of the final-state mesons,
respectively, in the $J/\psi(\psi^\prime)$ rest frame. We adopt
$\beta = 0.5\mbox {GeV}$, which is the same as
Refs.~\cite{amsler-close,close-kirk,close-zhao-scalar}. At leading
order the decays of $J/\psi(\psi^\prime)\to VP$ are via $P$-wave,
i.e. $l=1$. This form factor accounts for the size effects from
the spatial wavefunctions of the initial and final-state mesons.

\subsection{Decay of $J/\psi(\psi^\prime) \to VP$ via
EM interaction}

\begin{figure}
\begin{center}
\epsfig{file=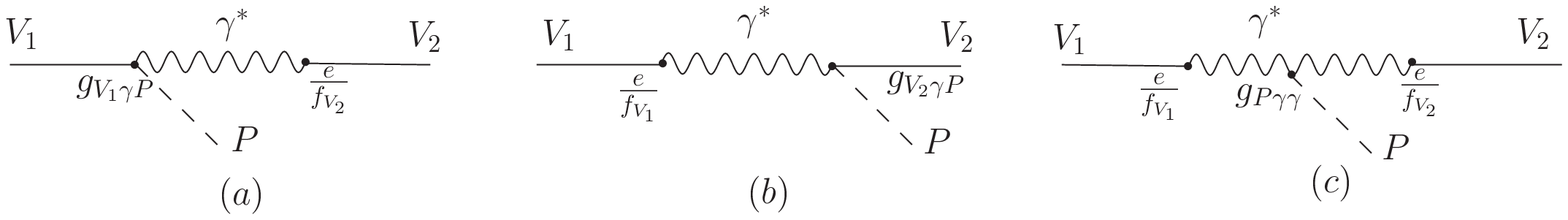, width=14cm,height=2.8cm} \caption{Schematic
diagrams for $J/\psi (\psi^\prime)\to \gamma^*\to VP$. }
\protect\label{fig-2}
\end{center}
\end{figure}

Detailed study of $J/\psi(\psi^\prime)\to\gamma^*\to VP$ in a VMD
model is presented in Ref.~\cite{zhao-li-chang}. In this process,
three independent transitions contribute to the total amplitude as
shown by Fig.~\ref{fig-2}. The advantage of treating this process
in the VMD model is to benefit from the available experimental
information for all those coupling vertices. As a result the only
parameter present in the EM transition amplitude is the form
factor for the virtual photon couplings to the initial ($J/\psi$
and $\psi^\prime$) and final state vector mesons ($\omega$,
$\phi$, $\rho$ and $K^*$). As shown in Ref.~\cite{zhao-li-chang},
the isospin violated channels, $\rho\eta$, $\rho\eta^\prime$,
$\omega\pi$ and $\phi\pi$, provides a good constraint on the
$\gamma^* VP$ form factors without interferences from the strong
decays. Therefore, a reliable estimate of the EM transitions can
be reached.

Following the analysis of Ref.~\cite{zhao-li-chang}, the invariant
transition amplitude for $J/\psi \to \gamma^* \to VP$ at tree
level can be expressed as:
\bea
{\cal M}_{EM} & \equiv & {\cal M}_A + {\cal M}_B + {\cal M}_C \nonumber\\
&=& \left( \frac{e}{f_{V2}}\frac{g_{V1\gamma P}}{M_{V1}}{\cal F}_a
+ \frac{e}{f_{V1}}\frac{g_{V2\gamma P}}{M_{V2}}{\cal F}_b +
\frac{e^2}{f_{V1} f_{V2}} \frac{g_{P\gamma\gamma}}{M_P}{\cal F}_c
\right)\epsilon_{\mu\nu\alpha\beta}\partial^\mu
V_1^\nu\partial^\alpha V_2^\beta P,
 \eea
where $e/f_V$ denotes the $\gamma^* V$ couplings, and $g_{V\gamma
P}$ is the coupling determined in the radiative decay of $V\to
\gamma V$; ${\cal F}_a$, ${\cal F}_b$ and ${\cal F}_c$ denote the
form factors corresponding to the transitions of Fig.~\ref{fig-2}.
For Fig.~\ref{fig-2}(a) and (b) a typical monopole (MP) form
factor is adopted: \be\label{mp} {\cal F}(q^2) = \frac
{1}{1-q^2/\Lambda^2} \ee with $\Lambda=0.542\pm 0.008$ GeV and
$\Lambda=0.577\pm 0.011 $ GeV determined with a constructive
(MP-C) or destructive phase (MP-D) between (a) and (b),
respectively. We think that the non-perturbative QCD effects might
have played a role in the transitions at $J/\psi$ energy. For
instance, in Fig.~\ref{fig-2}(a) and (b) a pair of quarks may be
created from vacuum as described by $^3P_0$ model, the pQCD
hadronic helicity-conservation due to the vector nature of gluon
is violated quite strongly. A monopole-like (MP) form factor
should be appropriate for coping with the suppression effects, and
it is consistent with the VMD framework. In principle such a form
factor can be tested experimentally via measuring the couplings of
the processes $J/\psi(\psi^\prime)\to P e^+ e^-$ and $e^+ e^-\to P
e^+ e^-$, respectively, when the integrated luminosity at $J/\psi$
and the suitable energies for $e^+e^-$ colliders is accumulated
enough.

Due to an error of missing a factor of $\sqrt{2}$ in Tab. I of
Ref.~\cite{zhao-li-chang}, we give here in Table~\ref{tab-gamma-V}
the coupling $e/f_V$ again. Also, we clarify that although the
overall factor of $\sqrt{2}$ will change the fitted cut-off energy
$\Lambda$ and the fitted branching ratios slightly in
Ref.~\cite{zhao-li-chang}, the pattern obtained there retains and
the major conclusion is intact.

The form factor for (c) has a form of:
\be
{\cal
F}_c(q_1^2,q_2^2)=\frac{1}{(1-q_1^2/\Lambda^2)(1-q_2^2/\Lambda^2)}
\ ,
\ee
where $q_1^2=M_{V1}^2$ and $q_2^2=M_{V2}^2$ are the squared
four-momenta carried by the time-like photons. We assume that the
$\Lambda$ is the same as in Eq.~(\ref{mp}), thus, ${\cal
F}_c={\cal F}_a {\cal F}_b$.

A parameter $\delta$ is introduced to take into account the
relative phase between the EM and strong transitions:
\be
{\cal M}={\cal M}_{3g} +e^{i\delta}{\cal M}_{EM}  \ .
\ee
It will then be determined by the experimental data in the
numerical fitting. In the limit of $\delta=0$, the relative phase
reduces to the same ones given in Ref.~\cite{seiden}.

\subsection{ Mixing of the $\eta-\eta^\prime$ and implication of a
$0^{-+}$ glueball }

The $\eta-\eta^\prime$ mixing is a long-standing question in the
literature. Here we would like to study the ``mixing" problem by
including empirically a possible glueball component in the $\eta$
and $\eta^\prime$ wavefunctions. We extend the mixing of the the
$\eta$ and $\eta^\prime$ as a consequence of the flavor singlet
$n\bar{n}$, $s\bar{s}$ and glueball mixing. The corresponding
glueball candidate is denoted as $\eta^{\prime\prime}$. In the
quark flavor basis, treating $\eta$, $\eta^\prime$ and
$\eta^{\prime\prime}$ as the eigenstates of the mass matrix $M$
with the eigenvalues of their masses $M_\eta$, $M_{\eta^\prime}$
and $M_{\eta^{\prime\prime}}$, we have
\be
U M U^{-1} = \left(\matrix{ M_\eta & 0 & 0 \cr 0 & M_{\eta^\prime}
& 0 \cr 0 & 0 & M_{\eta^{\prime\prime}} }\right) ,
\ee
where $U$ is the state mixing matrix:
 \be
 \left (\matrix{\eta \cr \eta^\prime \cr
\eta^{\prime\prime}}\right )= U  \left (\matrix{n\bar {n} \cr
s{\bar s} \cr G}\right )=
 \left (\matrix {x_1 & y_1 & z_1 \cr x_2 &y_2 & z_2 \cr x_3 & y_3 & z_3 }\right)
 \left (\matrix{n\bar {n} \cr s{\bar s} \cr
G}\right ) . \label{mix-matrix}
\ee

Three mixing schemes are applied to determine the mixing matrix
elememts.

I) CKM approach

By assuming that the $n\bar{n}$, $s\bar{s}$ and glueball $G$ make
a complete set of eigenstates, and physical states are their
linear combinations, we can express the mixing in the same way as
the CKM matrix with the phase $\delta=0$ (no $CP$ violation is
involved):
\be
U = \left (\matrix{ c_{12} c_{13} & s_{12} c_{13} & s_{13} \cr
-s_{13} c_{23}-c_{12} s_{23} s_{13} & c_{12} c_{23} -s_{12} s_{23}
s_{13} & s_{23} c_{13} \cr s_{12} s_{23} -c_{12} c_{23} s_{13} &
-c_{12}s_{23}-s_{12}c_{23}s_{13} & c_{23} c_{13} }\right) \ ,
\ee
where $c_{ij}\equiv \cos\theta_{ij}$ and $s_{ij}\equiv
\sin\theta_{ij}$ with $\theta_{ij}$ the mixing angles to be
determined by the experimental data. The feature of this approach
is that the completeness guarantees the unitary and orthogonal
relations of the matrix. However, it also implies that mixings
beyond $q\bar{q}$ and glueball are not allowed. This may be a
strong assumption since resonances and exotic states such as
tetraquarks with the same quantum number could also mix with the
$q\bar{q}$ and glueball. Because of this, the CKM approach will
test the extreme condition that only ground state $q\bar{q}$ and
$G$ mix with each other.

II) $q\bar{q}$-$G$ mixing due to higher Fock state $|gg\rangle$

This scenario is initiated by the QCD $U(1)_A$ anomaly. In a
series of studies by Feldmann {\it et al.}~\cite{fks,kroll-03}, it
is pointed out that an appropriate treatment of the mixing
requires to distinguish matrix elements of $\eta$ and
$\eta^\prime$ with local currents and overall state mixings.
Nonetheless, the divergences of axial-vector currents including
the axial anomaly connect the short-distance properties, i.e.
decay constants, with the long-distance phenomena such as the
mass-mixing, and highlight the twist-4 $|gg\rangle$ component in
the Fock state decomposition. In the quark flavor basis, higher
Fock state due to anomaly can give rise to a non-vanishing
$q\bar{q}\to |gg\rangle$ transition. Similar to the prescription
of Ref.~\cite{fks,kroll-03}, we introduce the glueball components
in $\eta$ and $\eta^\prime$ via the higher Fock state
decompositions in $\eta_n$ and $\eta_s$:
\bea\label{fock-states}
|\eta_n \rangle& =& \Psi_q|n\bar n\rangle
+\Psi_q^g|gg\rangle+\cdots, \nonumber \\
|\eta_s\rangle &=& \Psi_s|s\bar s\rangle
+\Psi_s^g|gg\rangle+\cdots ,
\eea
where $\Psi_q$ and $\Psi_s$ are amplitudes of the corresponding
$q\bar{q}$ while $\Psi_q^g$ and $\Psi_s^g$ are those of the
$|gg\rangle$ components; The dots denote higher Fock states with
additional gluon and/or $q\bar{q}$ which is beyond the
applicability of this approach. The presence of the gluonic Fock
states in association with isoscalar $n\bar{n}$ and $s\bar{s}$
breaks the orthogonality between $n\bar{n}$ and $s\bar{s}$, and we
parametrize such an effect in $\eta$ and $\eta^\prime$
wavefunctions:
\bea\label{mix-eta}
 |\eta\rangle &= & {1\over \sqrt {N_1}}[a(\cos{\phi}|n\bar
n\rangle-\sin {\phi}|s\bar s \rangle ) +b(\cos {\phi}-\sin
{\phi})|gg\rangle] , \nonumber\\
|\eta^\prime\rangle & = &  {1\over \sqrt {N_2}}[a(\sin{\phi}|n\bar
n\rangle+\cos {\phi}|s\bar s \rangle ) +b(\sin{\phi}+\cos
{\phi})|gg\rangle],
 \eea
where the normalization factors are $N_1=a^2+b^2(1-\sin {2\phi})$
and $N_2=a^2+b^2(1+\sin {2\phi})$, and parameters $a$ and $b$ are to
be determined by experiment. In this treatment, $a$ and $b$ now
correlate with the mixing angle $\phi\equiv \theta+\arctan\sqrt{2}$,
with $\theta$ as the octet-singlet mixing angle defined in the SU(3)
symmetry limit ($b\to 0$). For the commonly accepted range
$\theta\simeq -24.6^\circ$ or $-11.5^\circ$ from the linear or
quadratic mass formulae, respectively~\cite{pdg2006}, the glueball
component in $\eta$ has a strength of
$b(\cos\phi-\sin\phi)/\sqrt{N_1}$, which is relatively suppressed in
comparison with that in $\eta^\prime$, i.e.
$b(\sin\phi+\cos\phi)/\sqrt{N_2}$, for $0^\circ <\phi <90^\circ$.
This naturally gives rise to the scenario addressed in
Ref.~\cite{fks,kroll-03}.

Based on the unitary and orthogonal relation for three mixed
states, we can derive:
\be\label{glue-content}
\left\{\begin{array}{ccc} x_3^2 & = & 1-(x_1^2+x_2^2) , \\
y_3^2 & = & 1-(y_1^2+y_2^2) ,  \\
z_3^2 & = & 1-(z_1^2+z_2^2) .
\end{array}\right.
\ee
which will provide information about the $q\bar{q}$ and glueball
components in $\eta^\prime$ though it is not necessary for the
unitary and orthogonal relation being satisfied. If other higher
Fock states which are negligible in $\eta$ and $\eta^\prime$ are
present in $\eta^{\prime\prime}$ with sizable amplitude, further
constraints on the mixing wavefunction for $\eta^{\prime\prime}$
will be needed.

III) Mixing in an old perturbation theory

Considering the mixing between quarkonia and glueball at leading
order of a perturbation potential, i.e. the transition strength
between two states are much smaller than the mass difference of
these two states, we can then express the physical states
as~\cite{amsler-close}
\bea
|\eta\rangle & = & \frac{1}{C_1}\left[ | n\bar{n}\rangle
+\frac{\sqrt{2} f_b}{M_{n\bar{n}}-M_{s\bar{s}}}|s\bar{s}\rangle +
\frac{\sqrt{2}f_a}{M_{n\bar{n}}-M_G}|G\rangle \right] \nonumber\\
|\eta^\prime\rangle & =& \frac{1}{C_2}\left[ \frac{\sqrt{2}
f_b}{M_{s\bar{s}}-M_{n\bar{n}}}| n\bar{n}\rangle +
|s\bar{s}\rangle
+ \frac{\sqrt{2}f_a}{M_{s\bar{s}}-M_G}|G\rangle \right] \nonumber\\
|\eta^{\prime\prime}\rangle & =& \frac{1}{C_3}\left[
\frac{\sqrt{2} f_a}{M_G-M_{n\bar{n}}}| n\bar{n}\rangle +
\frac{\sqrt{2}f_a}{M_G-M_{s\bar{s}}}|s\bar{s}\rangle
+|G\rangle\right],
\eea
where $f_a\equiv \langle s\bar{s}| V_g |G\rangle =\langle n\bar{n}
| V_g |G\rangle/\sqrt{2} $ is the mixing strength for
glueball-$q\bar{q}$ transitions, while $f_b\equiv \langle
q\bar{q}| V_q | s\bar{s}\rangle $ is the $s\bar{s}$ and
non-strange $q\bar{q}$ mixing strength via transition potential
$V_q$. $C_{1,2,3}$ are the normalization factors. $M_{n\bar{n}}$,
$M_{s\bar{s}}$ and $M_G$ are masses for the pure $0^{-+}$
quarkonia and glueball states, respectively. They will be
determined with parameters $f_a$ and $f_b$ by fitting the
experimental data to satisfy the the physical masses for $\eta$,
$\eta^\prime$ and $\eta^{\prime\prime}$, and unitary and
orthogonal relations.

The above three parametrization schemes address different aspects
correlated with the glueball-$q\bar{q}$ mixing. The CKM approach
automatically satisfies the unitary and orthogonal relations, and
allows the mixing exclusively among those three states. In
reality, this may not be the case, and other configurations with
the same quantum number may also be present in the wavefunctions.
The second scheme introducing quarkonia-glueball mixing within
$\eta$ and $\eta^\prime$ via higher Fock state. In principle,
there is no constraint on the $\eta^{\prime\prime}$ configuration.
Thus, unitary and orthogonal relations do not necessarily apply to
those three states. The third scheme applies the old perturbation
theory and considers the non-vanishing transitions between those
pure states. As a result, a physical state will be a mixture of
quarkonia-glueball configurations. The unitary and orthogonal
relations will be a constraint in the determination of the
parameters. Comparing these three different parametrization
schemes, we expect that the quarkonia-glueball mixing mechanism
and implication of the pseudoscalar glueball candidate can be
highlighted.

In Table~\ref{tab-1}, the transition amplitudes for the strong
decays of $J/\psi(\psi^\prime)\to VP$ are given.

\section{Numerical results}

With the above preparations, we do the numerical calculations and
present the results in this section.

\subsection{Parameters and fitting scheme}

The parameters introduced in these three different approaches can
be classified into two classes. One consists of parameters which
are commonly defined in all three schemes, such as
$g_{J/\psi(\psi^\prime)}$, $r$ and $R$, and $\delta$. Parameter
$g_{J/\psi(\psi^\prime)}$ is the basic transition strength for
$J/\psi(\psi^\prime)\to VP$, and proportional to the wavefunctions
at origin for $J/\psi(\psi^\prime)$. Obviously, it has different
values for $J/\psi$ and $\psi^\prime$ decays, respectively.
Parameter $R$ and $r$ are the relative strengths of the SU(3)
flavor symmetry breaking and DOZI violation processes to
$g_{J/\psi(\psi^\prime)}$. They can also have different values in
$J/\psi$ and $\psi^\prime$ decays. Parameter $\delta$ indicates
the relative phases between the strong and EM transition
amplitudes.

The other class consists of parameters for quarkonia-glueball
mixings, for instance, the mixing angles in the CKM approach, and
masses $M_{n\bar{n}}$, $M_{s\bar{s}}$, and $M_G$ for the $0^{-+}$
quarkonia and glueball, respectively. These parameters depend on
the mixing schemes, and as mentioned earlier they give rise to
different scenarios concerning the $\eta$ and $\eta^\prime$
mixings. We shall discuss their properties in association with the
numerical results in the next section.

It should be noted that in the VMD model for the EM decay
transitions~\cite{zhao-li-chang} all couplings are determined
independently by accommodating the experimental information for
$V_1(V_2)\to e^+ e^-$, $P\to \gamma\gamma$, and $V_1(V_2)\to
\gamma P$ or $P\to \gamma V_2$, where $V_1=J/\psi$ or
$\psi^\prime$, $V_2=\omega$, $\phi$, $\rho$, or $K^*$, and
$P=\pi$, $\eta$, $\eta^\prime$ or $K$. This is essential for
accounting for the EM contributions properly. Meanwhile, the
radiative decays, such as $V\to\gamma \eta$ and $\gamma
\eta^\prime$, $\eta\to \gamma\gamma$ and $\eta^\prime\to
\gamma\gamma$, can probe the $q\bar{q}$ structure of the vector
and pseudoscalar mesons. Their constraints on the $\eta$ and
$\eta^\prime$ mixing have been embedded in constraining the EM
transitions. As shown in Ref.~\cite{zhao-li-chang}, all the
experimental data for $V_1(V_2)\to e^+ e^-$, $P\to \gamma\gamma$,
and $V_1(V_2)\to \gamma P$ or $P\to \gamma V_2$ have been
included.

Given a reliable description of the EM transitions, we can then
proceed to determine the strong transition parameters and
configuration mixings within the pseudoscalars by fitting  the
data for $J/\psi(\psi^\prime)\to VP$. The numerical study step is
taken as follows: i) Fit the data for $J/\psi(\psi^\prime)\to VP$
with only the strong decay transitions; ii) Fit the data for
$J/\psi(\psi^\prime)\to VP$ including the EM decay contributions.
Comparing those two situations, information about the role played
by the EM transitions, and their correlations with the strong
decay amplitudes can thus be extracted. The $\eta$ and
$\eta^\prime$ configurations can also be constrained. We emphasize
that a reasonable estimate of the EM contributions is a
prerequisite for a better understanding of the underlying
mechanisms in $J/\psi(\psi^\prime)\to VP$.

\subsection{Numerical results and analysis}

For each of these three parametrization schemes, three cases are
examined: i) with exclusive contributions from
$J/\psi(\psi^\prime)$ strong decays; ii) with a MP-D form factor
for the EM transitions; and iii) with a MP-C form factor for the
EM transitions. The parameters are listed in
Table~\ref{tab-2}-\ref{tab-4}, and the fitting results are listed
in Tables~\ref{tab-6} and \ref{tab-7} for $J/\psi$ and
$\psi^\prime$, respectively.

In general, without the EM contributions the fitting results have
a relatively larger $\chi^2$ value. With the EM contributions, the
results are much improved for both MP-D and MP-C model. To be more
specific, we first make an analysis of the commonly defined
parameters, i.e. $g_{J/\psi(\psi^\prime)}$, $r$, $R$ and $\delta$,
and then discuss the quarkonia-glueball mixing parameters for each
scheme.

For those commonly defined parameters their fitted values turn to
be consistent with each other in those three schemes. It is
interesting to compare the fitted values for the basic transition
amplitude $g_{J/\psi(\psi^\prime)}$ for $J/\psi$ and
$\psi^\prime$. It shows that the inclusion of the EM contributions
will bring significant changes to this quantity in $\psi^\prime$
decays, while it keeps rather stable in $J/\psi$ decays. This
agrees with the observation that the EM contributions in $J/\psi$
hadronic decays are less significant relative to the strong
ones~\cite{seiden,li-yu-fang}. The absolute value of
$g_{J/\psi(\psi^\prime)}$ for $\psi^\prime$ is naturally smaller
than that for $J/\psi$. Since $g_{J/\psi(\psi^\prime)}$ is
proportional to the wavefunctions at origins~\cite{zhao-li-chang},
the small fraction, $[g_{\psi^\prime}/g_{J/\psi}]^2\simeq 0.018$,
implies an overall suppression of the $\psi^\prime\to 3g$ form
factors in $\psi^\prime\to VP$, which is smaller than the pQCD
expectation, $\sim 12\%$. It is worth noting that this suppression
occurs not only to $\rho\pi$, but also to all the other $VP$
channels. We shall come back to this point in the later part.

Parameter $r$, denoting the OZI-rule violation effects, turns to
be sizeable in $J/\psi$ decays but smaller in $\psi^\prime$ decays
though relatively large uncertainties are accompanying. This is
consistent with our expectation that the DOZI processes in
$\psi^\prime$ decays will be relatively suppressed in comparison
with those in $J/\psi$ decays. However, it is noticed that $r$ has
quite large uncertainties in $\psi^\prime$ decays though the
central values are small. This reflects that data for
$\psi^\prime$ decays still possess relatively large errors, and
increased statistics may better constrain this parameter.

The SU(3) breaking parameter exhibits an overall consistency in the
fittings. Relatively large SU(3) flavor symmetry breaking turns to
occur in the $J/\psi$ decays compared with that in $\psi^\prime$. In
the case without EM contributions, the SU(3) breaking in
$\psi^\prime$ decays also turns to be large. This may reflect the
necessity of including the EM contributions. It shows that the SU(3)
symmetry breaking in $J/\psi$ can be as large as about $34\%$, while
it is about 5-20\% in $\psi^\prime$ decays.

The phase angle $\delta$ is fitted for $J/\psi$ and $\psi^\prime$,
respectively, when the EM contributions are included. It shows
that $J/\psi$ favors a complex amplitude introduced by the EM
transitions. In contrast, the strong and EM amplitudes in
$\psi^\prime$ decays is approximately out of phase. This means
that the strong and EM transitions will have destructive
cancellations in $\rho\pi$, $K^{*+}K^-+c.c.$, but constructively
interfere with each other in $K^{*0}\bar{K^0}+c.c.$ As
qualitatively discussed in Ref.~\cite{zhao-li-chang}, this phase
can lead to further suppression to $\psi^\prime\to \rho\pi$, and
also explain the difference between $K^{*+}K^-+c.c.$ and
$K^{*0}\bar{K^0}+c.c.$ The relative phases are in agreement with
the results of Ref.~\cite{suzuki}, but different from those in
Ref.~\cite{wym}.

\subsubsection{CKM approach}

In this mixing scheme it shows that mixing angles $\theta_{12}$ and
$\theta_{23}$ are well constrained by the experimental data while
large uncertainties occur to $\theta_{13}$ when the EM transitions
are included. The corresponding matrix element $\sin\theta_{12}$
favors a small value which implies small glueball component inside
$\eta$ meson. The mixed wavefunctions obtained with different EM
interferences are: \be\label{ckm-no-em} \mbox{without EM : \ \ \ }
\left\{
\begin{array}{ccl} |\eta\rangle &= & 0.915|n\bar{n}\rangle -0.404
|s\bar{s}\rangle +
0.003|G\rangle \\
|\eta^\prime\rangle & = & -0.398 |n\bar{n}\rangle -0.903
|s\bar{s}\rangle -0.163|G\rangle \\
|\eta^{\prime\prime}\rangle & = & 0.068 |n\bar{n}\rangle + 0.148
|s\bar{s}\rangle -0.987|G\rangle \
\end{array}\right. \ ;
\ee

\be\label{ckm-mp-d}
\mbox{MP-D Model: \ \ \ } \left\{ \begin{array}{ccl} |\eta\rangle
 & = & 0.904 |n\bar{n}\rangle -0.427 |s\bar{s}\rangle
+ 1.44\times 10^{-5}|G\rangle \\
|\eta^\prime\rangle & = & -0.421 |n\bar{n}\rangle - 0.892
|s\bar{s}\rangle -0.166|G\rangle \\
|\eta^{\prime\prime}\rangle & = & 0.071 |n\bar{n}\rangle + 0.150
|s\bar{s}\rangle -0.986|G\rangle \
\end{array}\right. \ ;
\ee
and
\be\label{ckm-mp-c} \mbox{MP-C Model: \ \ \ } \left\{
\begin{array}{ccl} |\eta\rangle & = & 0.901 |n\bar{n}\rangle
-0.433
|s\bar{s}\rangle
+5.0\times 10^{-7}|G\rangle \\
|\eta^\prime\rangle & = & -0.427 |n\bar{n}\rangle - 0.888
|s\bar{s}\rangle - 0.168|G\rangle \\
 |\eta^{\prime\prime}\rangle & = & 0.073
|n\bar{n}\rangle + 0.151 |s\bar{s}\rangle - 0.986|G\rangle \
\end{array}\right. \ .
\ee It shows that though $\theta_{13}$ has large uncertainties,
its central values indicate a small glueball component in $\eta$.
In particular, we find $\sin\theta_{13}\simeq 0$ in the MP-D
model, while $\sin\theta_{13} = 0.003$ or $5.0\times 10^{-7}$ from
the calculations without EM contributions or in the MP-C model,
respectively. In contrast, the amplitude of the glueball component
in the $\eta^\prime$ wavefunction is much larger. In all three
models (and also in all three mixing schemes), a stable glueball
mixing magnitude of $\sim 17\%$ is favored.

The automatically satisfied unitary and orthogonal conditions lead
to the prediction of the structure of the $\eta^{\prime\prime}$.
Assuming $\eta(1405)$ corresponding to $\eta^{\prime\prime}$, the
CKM scheme leads to a prediction of glueball-dominance inside
$\eta(1405)$. About $15\%$ of $s\bar{s}$ and $7\%$ of $n\bar{n}$
are also required in $\eta^{\prime\prime}$ and they favor to be
out of phase to the $G$ component.

\subsubsection{Quarkonia-glueball mixing due to higher Fock state}

Table~\ref{tab-3} shows that the octet-singlet mixing angle $\theta$
for $\eta$ and $\eta^\prime$ is within the reasonable range of
$-24.6^\circ\sim -11.5^\circ$ as found by other
studies~\cite{pdg2006}. Both $\eta$ and $\eta^\prime$ can
accommodate a small glueball component in association with the
dominant $q\bar{q}$. Parameter $a$ and $b$ are fitted by quite
different values with or without EM contributions. But remember that
it is the quantities $a/\sqrt{N_1}$, $b/\sqrt{N_1}$, $a/\sqrt{N_2}$
and $b/\sqrt{N_2}$ that alter the simple quark flavor mixing angles,
we should compare the mixed wavefunctions for $\eta$ and
$\eta^\prime$ in those fittings: \be \mbox{without EM : \ \ \ }
\left\{
\begin{array}{c} |\eta\rangle = 0.866 |n\bar{n}\rangle -0.494
|s\bar{s}\rangle -
0.070|G\rangle \\
|\eta^\prime\rangle = 0.480 |n\bar{n}\rangle + 0.841
|s\bar{s}\rangle +0.250|G\rangle \
\end{array}\right. \ ;
\ee

\be
\mbox{MP-D Model: \ \ \ } \left\{ \begin{array}{c} |\eta\rangle =
0.859 |n\bar{n}\rangle -0.510 |s\bar{s}\rangle
+ 0.046|G\rangle \\
|\eta^\prime\rangle = 0.503 |n\bar{n}\rangle + 0.847
|s\bar{s}\rangle -0.173|G\rangle \
\end{array}\right. \ ;
\ee
and
\be\label{mp-c-fock}
\mbox{MP-C Model: \ \ \ } \left\{ \begin{array}{c} |\eta\rangle =
0.859 |n\bar{n}\rangle -0.510 |s\bar{s}\rangle +
0.042|G\rangle \\
|\eta^\prime\rangle = 0.504 |n\bar{n}\rangle + 0.848
|s\bar{s}\rangle -0.163|G\rangle \
\end{array}\right. \ .
\ee Generally, it shows that the glueball component in $\eta$ is
small and in $\eta^\prime$ is relatively large. The $q\bar{q}$
contents are the dominant ones in their wavefunctions. In the case
without EM contributions, the signs for the glueball component are
opposite to those with the EM contributions included. Since the
glueball components are small, the $q\bar{q}$ mixing patterns are
quite similar in these three fittings, in particular, MP-D and MP-C
give almost the same results. We note in advance that the best
fitting results are obtained in the MP-C model. Thus, we concentrate
on the MP-C model here and try to extract information about the
glueball candidate $\eta^{\prime\prime}$. Similar discussions can be
applied to the other two schemes.


The above three equations indicate that orthogonality  between
$|\eta\rangle$ and $\eta^\prime\rangle$ is approximately
satisfied: $\langle\eta|\eta^\prime\rangle=-0.6\%$. This would
allow us to derive the mixing matrix elements for
$\eta^{\prime\prime}$ based on the unitary and orthogonal
relation: \be \left(\matrix {\eta \cr \eta^\prime \cr
\eta^{\prime\prime} }\right) = \left(\matrix {x_1 & y_1 & z_1 \cr
x_2 &y_2 & z_2 \cr x_3 & y_3 & z_3 }\right) \left(\matrix
{n\bar{n} \cr s\bar{s} \cr G}\right) = \left(\matrix {0.859 &
-0.510 & 0.042 \cr 0.504 & 0.848 & -0.163 \cr 0.090 & 0.140 &
0.986 }\right) \left(\matrix {n\bar{n} \cr s\bar{s} \cr G}\right)
, \label{mix-value} \ee where the orthogonal relation is satisfied
within 5\% of uncertainties. Note that an overall sign for
$|\eta^{\prime\prime}\rangle$ is possible.

So far, there is no much reliable information about the masses for
$|n\bar{n}\rangle$ and $|s\bar{s}\rangle$. Also, there is no firm
evidence for a $0^{-+}$ state (denoted as $\eta^{\prime\prime}$)
as a glueball candidate and mixing with $\eta$ and $\eta^\prime$.
Interestingly, the above mixing suggests a small $q\bar{q}$-$G$
coupling in the pseudoscalar sector, which is consistent with QCD
sum rule studies~\cite{narison}. As a test of this mixing pattern,
we can substitute the physical masses for $\eta$, $\eta^\prime$
and $0^{-+}$ resonances such as $\eta(1405)$, $\eta(1475)$ and
$\eta(1835)$~\cite{bes-x1835} into Eq.~(\ref{mix-value}) to derive
the pure glueball mass $M_G$, and we obtain, $M_G\simeq
M_{\eta^{\prime\prime}}$ due to the dominant of the glueball
component in $\eta^{\prime\prime}$. Empirically, this allows a
$0^{-+}$ glueball with much lighter masses than the prediction of
LQCD~\cite{lqcd}.

The amplitude for $J/\psi(\psi^\prime)\to \phi\eta^{\prime\prime}$
and $\omega\eta^{\prime\prime}$ can be expressed as \be
M_{\phi\eta^{\prime\prime}} = g_{J/\psi(\psi^\prime)} R [r x_3 +
(1+r) y_3 + z_3]{\cal F}({\bf p}) \ , \ee and \be
M_{\omega\eta^{\prime\prime}} = g_{J/\psi(\psi^\prime)} [(1+2r)
x_3 +\sqrt{2} R r y_3 +\sqrt{2} z_3]{\cal F}({\bf p}) \ .
\ee For those $0^{-+}$ resonances assumed to be the
$\eta^{\prime\prime}$ in Eq.~(\ref{mix-value}), predictions for
their production rates in $J/\psi(\psi^\prime)$ decays are listed
in Table~\ref{tab-5}. It shows that if those states are glueball
candidates, their branching ratios are likely at order of
$10^{-3}$ in $J/\psi\to \phi\eta^{\prime\prime}$ and
$\omega\eta^{\prime\prime}$, and at $10^{-5}$ in $\psi^\prime$
decays. We do not present the same calculations for the other two
mixing schemes since they all produce similar results.

Experimental signals for $\eta(1405)$ in $J/\psi$ radiative decays
were seen at Mark III~\cite{markIII-eta1405,bolton92b} and
DM2~\cite{augustin92}. In $J/\psi$ hadronic decays DM2 reported
$BR < 2.5\times 10^{-4}$ at $CL=90\%$ with the unknown
$\eta(1405)\to K\bar{K}\pi$ branching ratio
included~\cite{falvard-88}, and the search in Mark
III~\cite{mark-III} gave $BR(J/\psi\to\phi\eta(1405)\to \phi
K\bar{K}\pi) < 1.2\times 10^{-4} \ (90\% C.L.)$ and $BR(J/\psi\to
\omega\eta(1405)\to\omega K\bar{K}\pi) = (6.8 \begin{array}{c}
+1.9 \\ -1.6\end{array}\pm 1.7)\times 10^{-4}$.

The estimate of its total width is still controversial and ranges
from tens to more than a hundred MeV in different decay
modes~\cite{pdg2006}. This at least suggests that
$BR(\eta(1405)\to K\bar{K}\pi)\simeq 10 \sim 50\%$, and leads to
$BR(J/\psi\to \phi\eta(1405))\to\phi K\bar{K}\pi\sim < (1.49 \sim
7.45) \times 10^{-4}$. This range seems to be consistent with the
data~\cite{mark-III,falvard-88}. The prediction, $BR(J/\psi\to
\omega\eta(1405))= 7.32\times 10^{-3}$ turns to be larger than
$\phi$ channel. A similar estimate gives, $BR(J/\psi\to
\omega\eta(1405)\to \omega K\bar{K}\pi)\simeq (0.73 \sim 3.66)
\times 10^{-3}$, and can also be regarded as in agreement with the
data from Mark III~\cite{mark-III}. A search for $\eta(1405)$ at
BES-III in $J/\psi$ and $\psi^\prime$ hadronic decays will be
helpful to clarify its property.

The same expectation can be applied to $\eta(1475)$ and
$\eta(1835)$. If they are dominated by glueball components, their
production rate will be at order of $10^{-3}$ as shown in
Table~\ref{tab-5}. $\eta(1475)$, as the higher mass $0^{-+}$ in
the $\eta(1440)$ bump, tends to favor decaying into $K\bar{K}\pi$
(also via $a_0(980)\pi$ and $K^*\bar{K}$)~\cite{augustin92}. As
observed in experiment that the partial width for $\eta(1475)\to
K\bar{K}\pi$ is about 87 MeV~\cite{pdg2006}, an estimated
branching ratio of $BR(\eta(1475)\to K\bar{K}\pi)=0.1\sim 0.5$
will lead to $BR(J/\psi\to\phi\eta(1475)\to\phi
K\bar{K}\pi)=(1.33\sim 6.65)\times 10^{-4}$ and
$BR(J/\psi\to\omega\eta(1475)\to\omega K\bar{K}\pi)=(6.81\sim
34.05)\times 10^{-4}$. This value is compatible with the
production of $\eta(1405)$, thus, should have been seen at Mark
III in both $\omega$ and $\phi$ channel. However, signals for
$\eta(1475)$ is only seen in $\phi$ channel at
$BR(J/\psi\to\phi\eta(1475)\to\phi K\bar{K}\pi)< 2.1 \times
10^{-4}$ ($90\% C.L.)$~\cite{mark-III}. Such an observation is
more consistent with the $\eta(1475)$ being an $s\bar{s}$-dominant
state as the radial excited state of
$\eta^\prime$~\cite{cfl,barnes97}. Its vanishing branching ratio
in $\omega$ channel can be naturally explained by the DOZI
suppressions. We also note that $\eta(1475)$'s presence in
$\gamma\gamma\to K\bar{K}\pi$~\cite{acciarri01g} certainly
enhances its assignment as a radial excited $s\bar{s}$ state in
analogy with $\eta(1295)$ as the radial excited
$n\bar{n}$~\cite{cfl,barnes97}.

The narrow resonance $X(1835)$ reported by BES in $J/\psi\to\gamma
X(1835)\to\gamma\pi^+\pi^-\eta^\prime$ is likely to have
$J^{PC}=0^{-+}$~\cite{bes-x1835}, and could be the same state
reported earlier in $J/\psi\to \gamma p\bar{p}$~\cite{bes-ppbar}.
Its branching ratio is reported to be $BR(J/\psi\to \gamma X(1835)
\to\gamma \pi^+\pi^-\eta^\prime)= (2.2\pm 0.4 \pm 0.4)\times
10^{-4}$. By assuming that it is a pseudoscalar glueball
candidate, its production branching ratios in $\phi \eta(1835)$
and $\omega\eta(1835)$ are predicted in Table~\ref{tab-5}. If
$\pi^+\pi^-\eta^\prime$ is the dominant decay channel, one would
expect to have a good chance to see it in $\omega\eta(1835)$
channel. Since it is unlikely that a glueball state decays
exclusively to $\pi^+\pi^-\eta^\prime$ (and possibly $p\bar{p}$),
the predicted branching ratio $BR(J/\psi\to \omega X
(1835))=3.66\times 10^{-3}$ has almost ruled out its being a
glueball candidate though a search for the $X(1835)$ in its
recoiling $\omega$ should still be interesting. A number of
explanations for the nature of the $X(1835)$ were proposed in the
literature. But we are not to go to any details here.

With the above experimental observation, our results turn to favor
the $\eta(1405)$ being a pseudoscalar glueball candidate.

\subsubsection{Mixing in an old perturbation theory}

In this scheme, the masses of the $n\bar{n}$, $s\bar{s}$ and
glueball $G$ are treated as parameters along with the
``perturbative" transition amplitudes $f_a$ and $f_b$. What we
refer to as ``perturbative" here is that both $f_a$ and $f_b$ have
values much smaller than the mass differences between those mixed
states. The fitted results in Table~\ref{tab-4} indeed satisfy
this requirement. Parameters $f_a$ is fitted to be around 76 MeV,
which suggests a rather small glueball component in $\eta$ and
$\eta^\prime$ wavefunctions. In contrast, the $n\bar{n}$ and
$s\bar{s}$ mixing strength is slightly larger, i.e. $f_b\simeq 94$
MeV. The fitted masses $M_{n\bar{n}}\simeq 0.658$ GeV and
$M_{s\bar{s}}\simeq 0.853$ GeV are located between the $\eta$ and
$\eta^\prime$, which is consistent with the expectation of the
quark-flavor mixing picture with a mixing angle
$\phi=\theta+\arctan\sqrt{2}\sim 34.7^\circ$. The fitted mass for
the glueball is about 1.4 GeV, which is determined by the
assumption that the $\eta(1405)$ is a pseudoscalar glueball
candidate. Since the transition amplitude $f_a$ is relatively
small, the mixing does not bring significant differences between
the pure and physical glueball masses.

The wavefunctions for $\eta$, $\eta^\prime$ and
$\eta^{\prime\prime}$ in the three different considerations of the
EM transitions are \be \mbox{without EM : \ \ \ } \left\{
\begin{array}{ccl} |\eta\rangle & = & 0.864|n\bar{n}\rangle -0.479
|s\bar{s}\rangle -
0.157|G\rangle \\
|\eta^\prime\rangle & =& 0.480 |n\bar{n}\rangle +0.864
|s\bar{s}\rangle -0.152|G\rangle \\
|\eta^{\prime\prime}\rangle & =& 0.176 |n\bar{n}\rangle + 0.170
|s\bar{s}\rangle +0.970|G\rangle \
\end{array}\right. \ ;
\ee

\be \mbox{MP-D Model: \ \ \ } \left\{ \begin{array}{ccl}
|\eta\rangle & = & 0.864 |n\bar{n}\rangle -0.478 |s\bar{s}\rangle
- 0.155|G\rangle \\
|\eta^\prime\rangle & = & 0.479 |n\bar{n}\rangle + 0.865
|s\bar{s}\rangle -0.150|G\rangle \\
|\eta^{\prime\prime}\rangle & =& 0.174 |n\bar{n}\rangle + 0.168
|s\bar{s}\rangle +0.970|G\rangle \
\end{array}\right. \ ;
\ee
and
\be\label{mp-c-perturb}
\mbox{MP-C Model: \ \ \ } \left\{
\begin{array}{ccl} |\eta\rangle & =& 0.868 |n\bar{n}\rangle -
0.473
|s\bar{s}\rangle
-0.153|G\rangle \\
|\eta^\prime\rangle & =&  0.473 |n\bar{n}\rangle + 0.869
|s\bar{s}\rangle - 0.147|G\rangle \\
 |\eta^{\prime\prime} \rangle & =& 0.171
|n\bar{n}\rangle + 0.165 |s\bar{s}\rangle + 0.971|G\rangle \
\end{array}\right. \ .
\ee

In comparison with the first two mixing schemes, this approach in
the framework of old perturbation theory produces a similar mixing
pattern as that in Scheme-II except that the glueball component in
the $\eta$ wavefunction is quite significant, e.g. as shown in
Eq.~(\ref{mp-c-perturb}).

\subsubsection{The branching ratios for $J/\psi(\psi^\prime)\to VP$ and violation of the ``12\% rule" }

The fitted branching ratios for $J/\psi(\psi^\prime)\to VP$ are
listed in Tables~\ref{tab-6} and \ref{tab-7}, and the results for
the isospin violated channels are also included as a comparison.
It shows that all these three parametrizations can reproduce the
data quite well though there are different features arising from
the fitted results.

One predominant feature is that in all the schemes, parameter
$g_{J/\psi(\psi^\prime)}$ is found to have overall consistent
values for both $J/\psi$ and $\psi^\prime$. As pointed out earlier
that the relatively small value of $g_{J/\psi(\psi^\prime)}$ for
$\psi^\prime$ suggests an overall suppression of the
$\psi^\prime\to 3g$ form factor. With the destructive
interferences from the relatively large EM contributions, the
branching ratios for $\psi^\prime\to \rho\pi$ are further
suppressed and this leads to the abnormally small branching ratio
fractions between the exclusive decays of $J/\psi$ and
$\psi^\prime$. Numerically, this explains why the $12\%$ rule is
violated in $J/\psi(\psi^\prime)\to \rho\pi$. Nonetheless, such a
mechanism is rather independent of the final state hadron, thus,
should be more generally recognized in other exclusive channels.
This turns to be true. For instance, the large branching ratio
difference between the charged and neutral $K^*\bar{K}+c.c.$
channels highlights the interferences from the EM
transitions~\cite{zhao-li-chang}, where the relative phases to the
strong amplitudes are consistent with the expectations for the
$\rho\pi$ channel~\cite{seiden}.

In line with the overall good agreement of the fitting results to
the data is an apparent deviation in $\psi^\prime\to
\omega\eta^\prime$ in Scheme-II and III. The numerical fitting in
Scheme-II gives $BR(\psi^\prime\to\omega\eta^\prime)=9.49\times
10^{-8}$ in contrast with the data $(3.2\begin{array}{c} +2.5 \\
-2.1
\end{array})\times 10^{-5}$~\cite{pdg2006}. The significant deviation
from the experimental central value is allowed by the associated
large errors. In fact, this channel bears almost the largest
uncertainties in the datum set. The small values from the
numerical fitting also reflect the importance of the EM
interferences. Note that in the fitting with only strong
transitions, the branching ratio for $\omega\eta^\prime$ has
already turned to be smaller than the data. With the EM
contributions, which are likely to interfere destructively, this
channel is further suppressed.

In contrast, the CKM mixing scheme is able to reproduce the
$\psi^\prime\to\omega \eta^\prime$ branching ratio. This is
because the $q\bar{q}$ and $G$ components are in phase in the
$\eta^\prime$ wavefunctions in
Eqs.~(\ref{ckm-no-em})-(\ref{ckm-mp-c}). Since the
$\psi^\prime\to\omega\eta^\prime$ decay shows large sensitivities
to the mixing scheme, it is extremely interesting to have more
precise data for this channel as a directly test of the mixing
schemes proposed here. As BES-II may not be able to do any better
on this than been published~\cite{bes-04b}, CLEO-c with a newly
taken 25 million $\psi^\prime$ events can presumably clarify
this~\cite{cleo-c}.

In Table~\ref{tab-8}, we present the branching ratio fractions $R$
for all the exclusive decay channels for the MP-C model. The
extracted ratios from experimental data are also listed. Apart
from those three channels, $\omega\eta^\prime$, $\rho\eta^\prime$
and $\phi\pi$, of which the data still have large uncertainties,
the overall agreement is actually quite well. It clearly shows
that the ``12\% rule" is badly violated in those exclusive decay
channels, and the transition amplitudes are no longer under
control of pQCD leading twist~\cite{brodsky-lepage-81}. The power
suppression due to the violation of the hadronic helicity
conservation in pQCD will also be contaminated by other processes
which are much non-perturbative, hence the pQCD-expected simple
rule does not hold anymore. Note that it is only for those isospin
violated channels with exclusive EM transition as leading
contributions, may this simple rule be partly
retained~\cite{feldmann}.

\section{Summary}

In this work, we revisit $J/\psi(\psi^\prime)\to VP$ in a
parametrization model for the charmonium strong decays and a VMD
model for the EM decay contributions. By explicitly defining the
SU(3) flavor symmetry breaking and DOZI violation parameters, we
obtain an overall good description of the present available data. It
shows that a reliable calculation of the EM contributions is
important for understanding the overall suppression of the
$\psi^\prime\to 3g$ form factors. Our calculations suggest that
$\rho\pi$ channel is not very much abnormal compared to other $VP$
channels, and similar phenomena appear in $K^*\bar{K}$ as well.
Meanwhile, we strongly urge an improved experimental measurement of
the $\psi^\prime\to\omega\eta^\prime$ as an additional evidence for
the EM interferences. Although it is not for this analysis to answer
why $\psi^\prime\to 3g$ is strongly suppressed, our results identify
the roles played by the strong and EM transitions in
$J/\psi(\psi^\prime)\to VP$, and provide some insights into the
long-standing ``$\rho\pi$ puzzle".

Since the EM contributions are independently constrained by the
available experimental data~\cite{pdg2006}, the parameters
determined for the pseudoscalar in the numerical study, in turn,
can be examined by those data. In particular, we find that $\eta$
and $\eta^\prime$ allow a small glue component, which can be
referred to the higher Fock state contributions due to the
$U(1)_A$ anomaly. This gives rise to the correlated scenario
between the octet-singlet mixing angle and the decay constants as
addressed by Feldmann {\it et al}~\cite{fks,kroll-03}.

We are also interested in the possibility of a higher glue-dominant
state as a $0^{-+}$ glueball candidate. Indeed, based on the fact
that only a comparatively small glueball component exists in $\eta$
and $\eta^\prime$, we find that a $0^{-+}$ glueball which mixes with
$\eta$ and $\eta^\prime$ is likely to have nearly pure glueball
configuration. Although the obtained mixing matrix cannot pin down
the mass for a glueball state, the glueball dominance suggests that
such a glueball candidate will have large production rate in both
$\phi\eta^{\prime\prime}$ and $\omega\eta^{\prime\prime}$ at
$10^{-3}$. This enhance the assignment that $\eta(1405)$ is the
$0^{-+}$ glueball candidate if no signals for $\eta(1475)$ and
$\eta(1835)$ appear simultaneously in their productions with $\phi$
and $\omega$ in $J/\psi(\psi^\prime)$ decays. High-statistics search
for their signals in $\phi$ and $\omega$ channel at BESIII will be
able to clarify these results.

\section*{Acknowledgement}

Useful discussions with F.E. Close, C.Z. Yuan and B.S. Zou are
acknowledged. This work is supported, in part, by the U.K. EPSRC
(Grant No. GR/S99433/01), National Natural Science Foundation of
China (Grant No.10547001, No.90403031, and No.10675131), and
Chinese Academy of Sciences (KJCX3-SYW-N2).


\begin{table}[ht]
\begin{tabular}{c|c|c|c}
\hline Coupling const. $e/f_V$ & Values ($\times 10^{-2}$) & Total
width of $V$
& $BR(V\to e^+ e^-)$  \\[1ex]
\hline $e/f_\rho$ & 6.05 & 146.4 MeV & $(4.70\pm
0.08)\times 10^{-5}$ \\[1ex]
$e/f_\omega$ & 1.78 & 8.49 MeV & $(7.18\pm
0.12)\times 10^{-5}$\\[1ex]
$e/f_\phi$ & 2.26  & 4.26 MeV & $(2.97\pm
0.04)\times 10^{-4}$ \\[1ex]
$e/f_{J/\psi}$ & 2.71 & $93.4$ keV & $(5.94\pm 0.06)\%$  \\[1ex]
$e/f_{\psi^\prime}$ & 1.65 & 337 keV & $(7.35\pm 0.18)\times
10^{-3}$ \\[1ex] \hline
\end{tabular}
\caption{ The coupling constant $e/f_V$ determined in $V\to e^+
e^-$.  The data for branching ratios are from
PDG2006~\cite{pdg2006}. } \label{tab-gamma-V}
\end{table}

\begin{table}[ht]
\begin{tabular}{c|c}
\hline Decay channels  & Transition amplitude ${\cal M}=({\cal
M}_1+{\cal M}_2+{\cal M}_3)$
\\[1ex]
\hline  $\phi\eta$  & $g_{J/\psi(\psi^\prime)}R[\sqrt {2}r x_1
+R(1+r) y_1+ z_1]{\cal{F}}({\bf P})$
\\[1ex]
\hline  $\phi\eta^\prime$ & $g_{J/\psi(\psi^\prime)} R[\sqrt 2 r
x_2 +
R (1+r) y_2 + z_2]{\cal{F}}({\bf P})$ \\[1ex]
\hline $\omega\eta$ &  $ g_{J/\psi(\psi^\prime)} [(1+2r)x_1+\sqrt
2Rr y_1 + \sqrt 2 z_1]{\cal{F}}({\bf P})$\\[1ex]
\hline $\omega\eta^\prime$  &  $ g_{J/\psi(\psi^\prime)}
[(1+2r)x_2+\sqrt
2R r y_2 + \sqrt 2 z_2]{\cal{F}}({\bf P})$\\[1ex]
\hline $\rho^0\pi^0$ &  $g_{J/\psi(\psi^\prime)} {\cal{F}}({\bf P})$ \\[1ex]
\hline $\rho^+\pi^-$ or $\rho^-\pi^+$  &  $g_{J/\psi(\psi^\prime)} {\cal{F}}({\bf P})$ \\[1ex]
\hline $ { K}^{* 0}\bar{K^0}$ or $\bar{K^{*0}} K^0$  & $g_{J/\psi(\psi^\prime)} R{\cal{F}}({\bf P})$ \\[1ex]
\hline $ K^{*+}K^-$  or $ K^{*-}K^+$ &  $g_{J/\psi(\psi^\prime)} R{\cal{F}}({\bf P})$  \\[1ex]
\hline
\end{tabular}
\caption{ General expressions for the transition amplitudes for
$J/\psi(\psi^\prime)\to VP$ via strong interactions.}
\label{tab-1}
\end{table}


\begin{table}
 \begin{tabular}{|c|c|c|c|c|c|c|}
 \hline
& \multicolumn{2}{c|}{without EM}  & \multicolumn{2}{c|}{MP-D}   & \multicolumn{2}{c|}{MP-C}\\
          \cline{2-7}
  & $J/\psi$ & $\psi^\prime$ & $J/\psi$ & $\psi^\prime$& $J/\psi$ & $\psi^\prime$ \\
 \hline
 $r$ & $-0.265\pm 0.023$ &$-0.185\pm 0.194$  & $-0.270\pm 0.024$
 & $-0.080\pm 0.155$ &$-0.273\pm 0.024$ & $-0.054\pm 0.162$ \\
 \hline
 R& $0.660\pm 0.023$& $1.400\pm 0.112$  & $0.667\pm 0.024$
 &$1.293\pm 0.163$ & $0.667\pm 0.024$ & $1.233\pm 0.238$ \\
 \hline
$g_{J/\psi(\psi^\prime)}(\times 10^{-3})$ & $18.94\pm 0.64$&
$1.66\pm 0.15$ & $18.62\pm 0.63$ & $2.06\pm 0.25$ &
$18.66\pm 0.63$ & $2.22 \pm 0.49$ \\
\hline $\delta$& $-$ & $-$ & $75.0^\circ\pm 5.2^\circ$ &
$143.9^\circ\pm 14.6^\circ$
& $73.7^\circ \pm 5.8^\circ$ & $159.5^\circ \pm 29.9^\circ$ \\
\hline $\theta_{12}$ & \multicolumn{2}{c|}{$156.2^\circ \pm
2.0^\circ$}
 & \multicolumn{2}{c|}{$154.7^\circ \pm
2.4^\circ$}
 & \multicolumn{2}{c|}{$154.3^\circ \pm
2.4^\circ$} \\
 \hline $\theta_{13}$ & \multicolumn{2}{c|}{$179.8^\circ \pm
0.7^\circ$}
 & \multicolumn{2}{c|}{$180.0^\circ \pm
0.3^\circ$}
 & \multicolumn{2}{c|}{$180.0^\circ \pm
0.2^\circ$} \\
 \hline $\theta_{23}$ & \multicolumn{2}{c|}{$9.4^\circ \pm
1.7^\circ$}
 & \multicolumn{2}{c|}{$9.5^\circ \pm
1.9^\circ$}
 & \multicolumn{2}{c|}{$9.7^\circ \pm
1.9^\circ$} \\
\hline $\chi^2$/d.o.f &\multicolumn{2}{c|}{37.0/9}& \multicolumn{2}{c|}{9.8/11}&\multicolumn{2}{c|}{9.1/11} \\
\hline
 \end{tabular}
\caption{Parameters introduced in the overall fitting of
$J/\psi(\psi^\prime) \to VP$ in Scheme-I, i.e. the CKM approach.
}\label{tab-2}
 \end{table}

\begin{table}
 \begin{tabular}{|c|c|c|c|c|c|c|}
 \hline
  & \multicolumn{2}{c|}{without EM}
          & \multicolumn{2}{c|}{MP-D} & \multicolumn{2}{c|}{MP-C}\\
          \cline{2-7}
  & $J/\psi$ & $\psi^\prime$ & $J/\psi$ & $\psi^\prime$& $J/\psi$ & $\psi^\prime$ \\
 \hline
 $r$ & $-0.183\pm 0.034$ &$0.101\pm 0.237$  & $-0.308\pm 0.037$
 & $-0.237\pm 0.155$ &$-0.307\pm 0.038$ & $-0.206\pm 0.129$ \\
 \hline
 R& $0.661\pm 0.024$& $1.400\pm 0.106$  & $0.672\pm 0.027$
 &$1.098\pm 0.232$ & $0.674\pm 0.028$ & $1.065\pm 0.181$ \\
 \hline
$g_{J/\psi(\psi^\prime)}(\times 10^{-3})$ & $18.91\pm 0.64$&
$1.63\pm 0.15$ & $18.45\pm 0.69$ & $2.40\pm 0.53$ &
$18.45\pm 0.70$ & $2.54 \pm 0.44$ \\
\hline $\delta$& $-$ & $-$ & $74.0^\circ\pm 5.2^\circ$ &
$30.7^\circ\pm 20.4^\circ$
& $72.3^\circ \pm 6.0^\circ$ & $11.0^\circ \pm 61.8^\circ$ \\
\hline $\theta$ & \multicolumn{2}{c|}{$-25.0^\circ \pm 3.7^\circ$}
 & \multicolumn{2}{c|}{$-24.0^\circ\pm 0.8^\circ$}
 & \multicolumn{2}{c|}{$-24.0^\circ \pm 0.9^\circ$} \\
\hline $a$ & \multicolumn{2}{c|}{$0.673\pm 0.159 $} &
\multicolumn{2}{c|}{$(3.15\pm 0.90)\times 10^{-2}$}
& \multicolumn{2}{c|}{$(3.44\pm 1.03)\times 10^{-2}$} \\
\hline $b$& \multicolumn{2}{c|}{$-0.127\pm 0.013$} &
\multicolumn{2}{c|}{$(4.03\pm 0.60)\times 10^{-3}$} & \multicolumn{2}{c|}{$(4.13\pm 0.65)\times 10^{-3}$}  \\
\hline $\chi^2$/d.o.f &\multicolumn{2}{c|}{41.1/9}& \multicolumn{2}{c|}{11.2/11}&\multicolumn{2}{c|}{10.7/11} \\
\hline
 \end{tabular}
\caption{Parameters introduced in the overall fitting of
$J/\psi(\psi^\prime) \to VP$ in Scheme-II, i.e. $q\bar{q}$-G
mixing due to higher Fock state contributions. }\label{tab-3}
 \end{table}


\begin{table}
 \begin{tabular}{|c|c|c|c|c|c|c|}
 \hline
  & \multicolumn{2}{c|}{without EM}
          & \multicolumn{2}{c|}{MP-D} & \multicolumn{2}{c|}{MP-C}\\
          \cline{2-7}
  & $J/\psi$ & $\psi^\prime$ & $J/\psi$ & $\psi^\prime$& $J/\psi$ & $\psi^\prime$ \\
 \hline
 $r$ & $-0.035\pm 0.056$ &$0.115\pm 0.193$  & $-0.037\pm 0.055$
 & $0.031\pm 0.151$ &$-0.044\pm 0.057$ & $0.041\pm 0.141$ \\
 \hline
 R& $0.698\pm 0.025$& $1.400\pm 0.297$  & $0.715\pm 0.027$
 &$1.079\pm 0.155$ & $0.718\pm 0.028$ & $1.047\pm 0.158$ \\
 \hline
$g_{J/\psi(\psi^\prime)}(\times 10^{-3})$ & $18.05\pm 0.64$&
$1.65\pm 0.15$ & $17.53\pm 0.64$ & $2.44\pm 0.34$ &
$17.52\pm 0.66$ & $2.57 \pm 0.42$ \\
\hline $\delta$& $-$ & $-$ & $74.9^\circ\pm 5.07^\circ$ &
$149.2^\circ\pm 19.6^\circ$
& $73.8^\circ \pm 5.8^\circ$ & $168.2^\circ \pm 136.7^\circ$ \\
\hline $M_{n\bar{n}}$ (GeV) & \multicolumn{2}{c|}{$0.658 \pm
0.083$}
 & \multicolumn{2}{c|}{$0.658\pm 0.055$}
 & \multicolumn{2}{c|}{$0.657\pm 0.081$} \\
 \hline $M_{s\bar{s}}$ (GeV) & \multicolumn{2}{c|}{$0.855 \pm 0.076$}
 & \multicolumn{2}{c|}{$0.853\pm 0.057$}
 & \multicolumn{2}{c|}{$0.853 \pm 0.077$} \\
 \hline $M_{G}$ (GeV) & \multicolumn{2}{c|}{$1.389 \pm 0.093$}
 & \multicolumn{2}{c|}{$1.390\pm 0.093$}
 & \multicolumn{2}{c|}{$1.391 \pm 0.095$} \\
\hline $f_a$ (GeV) & \multicolumn{2}{c|}{$0.077 \pm 0.039$}
 & \multicolumn{2}{c|}{$0.076\pm 0.022$}
 & \multicolumn{2}{c|}{$0.076 \pm 0.038$} \\
\hline $f_b$ (GeV)& \multicolumn{2}{c|}{$0.094\pm 0.033 $} &
\multicolumn{2}{c|}{$0.093\pm 0.034$}
& \multicolumn{2}{c|}{$0.091\pm 0.036$} \\
\hline $\chi^2$/d.o.f &\multicolumn{2}{c|}{46.3/13}& \multicolumn{2}{c|}{18.9/13}&\multicolumn{2}{c|}{18.3/13} \\
\hline
 \end{tabular}
\caption{Parameters introduced in the overall fitting of
$J/\psi(\psi^\prime) \to VP$ in Scheme-III, i.e. mixing via an old
perturbation theory. }\label{tab-4}
 \end{table}

\begin{table}
 \begin{tabular}{|c|c|c|c|c|}
 \hline
 & \multicolumn{2}{c|}{$BR(J/\psi\to V\eta^{\prime\prime}) \ (\times 10^{-3})$}
 & \multicolumn{2}{c|}{$BR(\psi^\prime\to V\eta^{\prime\prime}) \ (\times 10^{-5})$}\\
          \cline{2-5}
 $ \eta^{\prime\prime} $ & $ \ \ \ \ \ \ \phi\eta^{\prime\prime} \ \ \ \ \ \ $
 & $  \omega\eta^{\prime\prime} $
 & $ \ \ \ \ \ \ \phi\eta^{\prime\prime} \ \ \ \ \ \ $ & $\omega\eta^{\prime\prime}$ \\
 \hline
 $\eta(1405)$ & 1.49 & 7.32  & 4.07 & 7.17 \\
 \hline
 $\eta(1475)$ & 1.33  &  6.81  & 3.96 & 7.03  \\
 \hline
$\eta(1835)$ & 0.44  & 3.66 & 3.09 & 5.93   \\
\hline
 \end{tabular}
\caption{Predictions for production of $\eta^{\prime\prime}$ in
$J/\psi(\psi^\prime)\to VP$  by assuming it is $\eta(1405)$,
$\eta(1475)$ and $\eta(1835)$, respectively, in Scheme-II.
Calculations by Scheme-I and III give similar results with the
same order of magnitude. }\label{tab-5}
 \end{table}

\begin{table}[ht]
\begin{tabular}{c|c|c|c|c|c|c}
\hline Decay channels &without EM  & Scheme-I & Scheme-II &
Scheme-II
 & Scheme-III  & Exp.
data  \\
 & & (MP-C) & (MP-D) & (MP-C) &(MP-C) & \\[1ex]\hline
$\rho^0\pi^0$ & $5.64\times 10^{-3}$ & $5.76\times 10^{-3}$ & $
5.64\times 10^{-3}$ & $5.65\times 10^{-3}$ &$5.09\times 10^{-3}$&
$(5.6\pm 0.7)\times 10^{-3}$ \\[1ex]
$\rho\pi$ & $ 1.69\times 10^{-2}$& $1.73\times 10^{-2}$ &
$1.70\times 10^{-2}$  & $1.69\times 10^{-2}$ &$1.53\times 10^{-2}$
& $(1.69\pm 0.15)\times 10^{-2}$
 \\[1ex]
$\omega\eta$ & $1.55\times 10^{-3}$ & $1.53\times 10^{-3}$&
$1.60\times 10^{-3}$  & $1.60\times 10^{-3}$ & $1.75\times
10^{-3}$ &
$(1.74\pm 0.20)\times 10^{-3}$ \\[1ex]
$\omega\eta^\prime$ & $1.82\times 10^{-4}$& $1.82\times 10^{-4}$&
$1.83\times 10^{-4}$ & $1.83\times 10^{-4}$ & $1.80\times 10^{-4}$
& $(1.82\pm 0.21)\times 10^{-4}$ \\[1ex]
$\phi\eta$ &  $6.45\times 10^{-4}$ & $6.40\times 10^{-4}$ &
$6.91\times 10^{-4}$ & $6.94\times 10^{-4}$  & $5.92\times
10^{-4}$ &
$(7.4\pm 0.8)\times 10^{-4}$  \\[1ex]
$\phi\eta^\prime$ & $4.00\times 10^{-4}$ & $4.22\times 10^{-4}$&
$4.00\times 10^{-4}$ & $4.00\times 10^{-4}$ & $3.17\times 10^{-4}$
&
$(4.0\pm 0.7)\times 10^{-4}$ \\[1ex]
$K^{*+}K^-+c.c.$ & $4.67\times 10^{-3}$ & $5.01\times 10^{-3}$&
$5.03\times 10^{-3}$ & $5.03\times 10^{-3}$ & $5.11\times 10^{-3}$
&
$(5.0\pm 0.4)\times 10^{-3}$ \\[1ex]
$K^{*0}\bar{K^0}+c.c.$ & $4.66\times 10^{-3}$ & $4.30\times
10^{-3}$& $4.25\times 10^{-3}$ & $4.23\times 10^{-3}$ &
$4.39\times 10^{-3}$
& $(4.2\pm 0.4)\times 10^{-3}$ \\[1ex]\hline
$\rho\eta$ & $-$ & $1.5\times 10^{-4}$& $1.1\times 10^{-4}$  &
\multicolumn{2}{c|}{$1.5\times 10^{-4}$} &
$(1.93\pm 0.23)\times 10^{-4}$ \\[1ex]
$\rho\eta^\prime$ & $-$  & $7.9\times 10^{-5}$ & $4.2\times
10^{-5}$ & \multicolumn{2}{c|}{$7.9\times 10^{-5}$} &
$(1.05\pm 0.18)\times 10^{-4}$ \\[1ex]
$\omega\pi$ & $-$  & $3.3\times 10^{-4}$& $4.2\times 10^{-4}$ &
\multicolumn{2}{c|}{$3.3\times 10^{-4}$} & $(4.5\pm 0.5)\times
10^{-4}$\\[1ex]
$\phi\pi$ & $-$ & $9.9\times 10^{-7}$ &  $8.0\times 10^{-7}$  &
\multicolumn{2}{c|}{$9.9\times 10^{-7}$}  & $<
6.4\times 10^{-6}$ \\[1ex]\hline
\end{tabular}
\caption{ Fitted branching ratios for $J/\psi\to VP$ in different
parametrization schemes for the isoscalar mixings. Column (without
EM) is for results without EM contributions; Columns of MP-C
correspond to processes Fig.~\ref{fig-2}(a) and (b) in a
constructive phase with an effective mass $\Lambda=0.616$ GeV; and
Column MP-D corresponds to (a) and (b) in a destructive phase with
$\Lambda=0.65$ GeV, which are the same as in
Ref.~\cite{zhao-li-chang}. The isospin violated channels (last four
channels) are also listed. The experimental branching ratios are
from PDG2006~\cite{pdg2006}. } \label{tab-6}
\end{table}

\begin{table}[ht]
\begin{tabular}{c|c|c|c|c|c|c}
\hline Decay channels & without EM  & Scheme-I & Scheme-II &
Scheme-II
 & Scheme-III  & Exp.
data  \\
 & & (MP-C) & (MP-D) & (MP-C) &(MP-C) & \\[1ex]\hline
$\rho^0\pi^0$ & $0.89\times 10^{-5}$ & $0.71 \times 10^{-5}$ &
$0.88\times 10^{-5}$ & $0.97 \times 10^{-5}$ & $1.02 \times
10^{-5}$ &
*** \\[1.5ex]
$\rho\pi$ & $2.68\times 10^{-5}$ & $2.03\times 10^{-5}$&
$2.58\times 10^{-5}$ & $2.78\times 10^{-5}$ & $2.92\times 10^{-5}$
&
$(3.2\pm 1.2)\times 10^{-5}$ \\[1.5ex]
$\omega\eta$ & $6.36\times 10^{-6}$ & $6.08\times 10^{-6}$&
$5.63\times 10^{-6}$ & $5.35\times 10^{-6}$ & $4.97\times 10^{-6}$
&
$< 1.1\times 10^{-5}$ \\[1ex]
$\omega\eta^\prime$ & $1.35\times 10^{-6}$ & $9.68\times 10^{-6}$&
$1.96\times 10^{-7}$ & $9.49\times 10^{-8}$ & $5.21\times 10^{-7}$
& $(3.2\begin{array}{c} +2.5\\ -2.1\end{array})\times 10^{-5}$\\[1.5ex]
$\phi\eta$ & $0.95\times 10^{-5}$ & $1.78\times 10^{-5}$&
$2.02\times 10^{-5}$ & $2.04\times 10^{-5}$ & $1.96\times 10^{-5}$
&
$(2.8\begin{array}{c} +1.0\\ -0.8\end{array})\times 10^{-5}$  \\[1.5ex]
$\phi\eta^\prime$ & $1.75\times 10^{-5}$ & $2.11\times 10^{-5}$&
$2.13\times 10^{-5}$ & $2.41\times 10^{-5}$ & $2.80\times 10^{-5}$
&
$(3.1\pm 1.6)\times 10^{-5}$  \\[1.5ex]
$K^{*+}K^-+c.c.$ & $3.50\times 10^{-5}$ & $2.28\times 10^{-5}$&
$2.07\times 10^{-5}$ & $2.01\times 10^{-5}$ & $2.00\times 10^{-5}$
&
$(1.7\begin{array}{c} +0.8\\ -0.7\end{array} )\times 10^{-5}$ \\[1.5ex]
$K^{*0}\bar{K^0}+c.c.$ & $3.50\times 10^{-5}$ & $1.18\times
10^{-4}$ & $1.18\times 10^{-4}$ & $1.17\times 10^{-4}$ &
$1.13\times 10^{-4}$
& $(1.09\pm 0.20)\times 10^{-4}$\\[1.5 ex]\hline
$\rho\eta$ & $-$  & $1.4\times 10^{-5}$& $9.4\times 10^{-6}$
&\multicolumn{2}{c|}{$1.4\times 10^{-5}$}  &
$(2.2\pm 0.6)\times 10^{-5}$ \\[1ex]
$\rho\eta^\prime$ & $-$  & $7.4\times 10^{-6}$& $3.9\times
10^{-6}$ &\multicolumn{2}{c|}{$7.4\times 10^{-6}$} &
$(1.9 \begin{array}{c} +1.7 \\ -1.2\end{array})\times 10^{-5}$ \\[1ex]
$\omega\pi$ & $-$  & $3.0\times 10^{-5}$& $3.9\times 10^{-5}$
&\multicolumn{2}{c|}{$3.0\times 10^{-5}$} & $(2.1\pm 0.6)\times
10^{-5}$ \\[1ex]
$\phi\pi$ & $-$  & $7.3\times 10^{-8}$& $9.6\times 10^{-8}$ &
\multicolumn{2}{c|}{$7.3\times 10^{-8}$}
& $<4\times 10^{-6}$ \\[1ex]\hline
\end{tabular}
\caption{ Fitted branching ratios for $\psi^\prime\to VP$. The
notations are the same as Table~\ref{tab-6}. The stars ``***" in
$\rho^0\pi^0$ channel denotes the unavailability of the data.}
\label{tab-7}
\end{table}


\begin{table}[ht]
\begin{tabular}{c|c|c|c|c}
\hline
 Decay channels & Scheme-I (\%) & Scheme-II (\%)  & Scheme-III (\%) & Exp. data (\%)
  \\[1ex]\hline
$\rho\pi$ & 0.12 & 0.15 & 0.19 & $0.2\pm 0.1$
 \\[1ex]
$\omega\eta$ & 0.40 & 0.35 & 0.28 & $< 0.6\pm 0.1$  \\[1ex]
$\omega\eta^\prime$ & 5.33 & 0.11 & 0.29 & $18.5\pm 13.2$  \\[1.ex]
$\phi\eta$ & 2.78 & 2.93 & 3.30 & $4.1\pm 1.6$  \\[1.ex]
$\phi\eta^\prime$ & 5.00 & 5.34 & 8.86 & $8.7\pm 5.5$ \\[1.ex]
$K^{*+}K^-+c.c.$ & 0.45 & 0.41 & 0.39 & $0.4\pm 0.2$ \\[1ex]
$K^{*0}\bar{K^0}+c.c.$ & 2.74 & 2.79 & 2.67 & $2.7\pm 0.7$  \\[1ex]
\hline $\rho\eta$ & \multicolumn{3}{c|}{8.97} &
$11.5\pm 5.0 $\\[1.ex]
$\rho\eta^\prime$  & \multicolumn{3}{c|}{9.44} &
$23.5\pm 17.8$ \\[1ex]
$\omega\pi$  & \multicolumn{3}{c|}{9.01} & $5.0\pm 1.8$\\[1ex]
$\phi\pi$ &  \multicolumn{3}{c|}{7.41} & $< 62.5 $ \\[1ex]\hline
\end{tabular}
\caption{ Branching ratio fractions for all $VP$ channels in the
MP-C model. The isospin violated channels are also included. }
\label{tab-8}
\end{table}
%


\begin{figure}[htbp]
\begin{center}
\epsfig{file=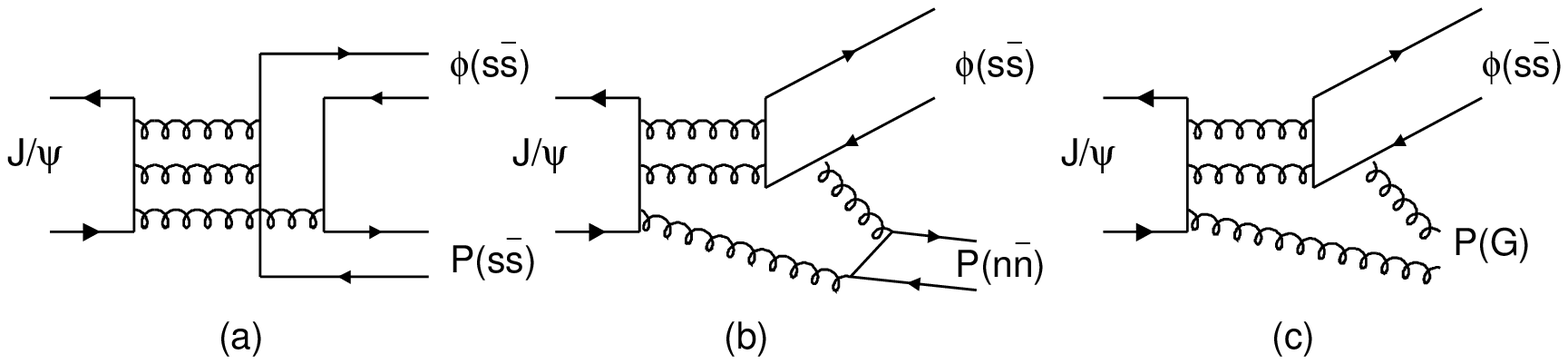,width=14.0cm,height=7cm} \caption{Schematic
diagrams for $J/\psi \to \phi P$ via strong interaction, where the
production of different components of the pseudoscalar $P$ is
demonstrated via (a): SOZI process; (b) DOZI process; and (c)
glueball production. Similar prcesses apply to other $VP$ channels
as described in the text. }
\end{center}
\label{fig-1}
\end{figure}

\end{document}